\documentclass[reprint,prd,aps,amsmath,superscriptaddress,twocolumn,amssymb,nofootinbib]{revtex4}

\usepackage{hyperref}
\usepackage{epsfig}
\usepackage{graphicx}

\newcommand{\bea}{\begin{eqnarray}}
\newcommand{\eea}{\end{eqnarray}}
\newcommand{\beq}{\begin{equation}}
\newcommand{\eeq}{\end{equation}}

\begin{document}

\title{The curvaton scenario within the MSSM and predictions for non-Gaussianity}

\author{Anupam Mazumdar}
\affiliation{Physics Department, Lancaster University, Lancaster LA1 4YB, UK}
\affiliation{Niels Bohr Institute, Copenhagen, Blegdamsvej-17, Denmark}
\author{Seshadri Nadathur}
\affiliation{Rudolf Peierls Centre for Theoretical Physics, University of Oxford, Oxford OX1 3NP, UK}


\begin{abstract}
We provide a model in which both the inflaton and the curvaton are obtained from within the minimal supersymmetric Standard Model, with known gauge and Yukawa interactions. Since now both the inflaton and curvaton fields are successfully embedded within the same sector, their decay products thermalize very quickly before the electroweak scale. This results in two important features of the model: firstly, there will be no residual isocurvature perturbations, and secondly, observable non-Gaussianities can be generated with the non-Gaussianity parameter $f_\mathrm{NL}\sim {\cal O}(5-1000)$ being determined solely by the combination of weak-scale physics and the Standard Model Yukawas.
\end{abstract}

\maketitle

The curvaton scenario~\cite{david,enqvist,moroi,DimopoulosLyth} is an alternative mechanism for the generation of the primordial perturbations whose spectrum is observed in the cosmic microwave background (CMB)~\cite{WMAP}. In this scenario, the density perturbations are sourced by the quantum fluctuations of a light scalar field $\phi$, the curvaton, which makes a negligible contribution to the energy density during inflation and decays after the decay of the inflaton field $\sigma$.~(For a review on inflation including the curvaton mechanism, see~\cite{RM}.) The advantage of the curvaton mechanism is that it can in principle generate measurable non-Gaussianity \cite{david,nG-curvaton}in the primordial density perturbations and also significant residual isocurvature perturbations, neither of which are possible in the usual single-field inflation models. Both signatures are detectable, and if either were to be observed, this would strongly favour the curvaton hypothesis.

If the curvaton does not completely dominate the energy density at the time of its decay, the process of conversion of initial isocurvature perturbations into adiabatic curvature perturbations can enhance non-Gaussian fluctuations to the level where they might be constrained by the Planck satellite. The enhancement in non-Gaussianity is given by $f_\mathrm{NL}\sim 5/(4r)$ for $r< 1$, where $r\equiv \rho_{\phi}/\rho_\mathrm{rad}$ at the time the curvaton decays~\cite{david}. Planck is expected to be able to detect non-Gaussianity of the order $f_\mathrm{NL}\gtrsim5$~\cite{Komatsu:2001rj}. To achieve detectable $f_\mathrm{NL}$ thus requires small $r$.

However, if either the curvaton or the inflaton belong to a hidden sector beyond the Standard Model (SM), they may decay into other fields beyond the SM degrees of freedom ({\it dof}). There is no guarantee that the hidden and visible sector {\it dof} should reach thermal equilibrium before Big Bang Nucleosynthesis (BBN)~\cite{BBN} takes place. In this case, residual isocurvature perturbations are expected to be in conflict with CMB data, which constrain them to be less than $10\%$~\cite{WMAP}. If the curvaton belongs to the visible sector but the inflaton does not, a value of $r\sim1$ would avoid this conflict \cite{Gordon:2002gv} but would render any non-Gaussianity undetectable. Note that if $r\sim1$ the curvaton is solely responsible for exciting all the SM {\it dof} so it must carry the SM charges~\cite{Curvaton-enqvist,Kasuya:2004}.

For the curvaton model to be observationally distinguishable, we wish the model to be able to create detectable non-Gaussianity. For this, $r$ must be small and both the inflaton and curvaton decay products must thermalize before the time of nucleosynthesis, as there are stringent constraints on any non-SM like hidden radiation after BBN~\cite{BBN}. In order to achieve this, we wish to place the entire inflaton-curvaton paradigm within a particle physics model where all the interactions are well constrained by the weak scale physics.

Recently, the inflationary paradigm has been embedded within the minimal supersymmetric Standard Model (MSSM) with known gauge interactions~\cite{AEGM,AKM}. The aim of this letter is to show, for the first time, that it is  possible to embed both the inflaton and curvaton within MSSM, without involving any hidden sector. We thus provide a solution to a general problem of the curvaton scenario, i.e., how to generate measurable non-Gaussianity without large residual isocurvature fluctuations.

Let us first consider the total potential to be the sum of inflaton vacuum energy, denoted by $V_{0}$, and curvaton potential $V(\phi)$
\begin{equation}
V_{total}=V_{0}+V({\phi})\,.
\end{equation}
We assume $V''(\phi)\sim m^{2}_{\phi}(\phi_I)\ll H^{2}_I\sim V_{0}/M_{\rm P}^{2}$ ($M_{\rm P}\sim 10^{18}$~GeV) where the subscript $I$ indicates the quantities are evaluated during inflation. This condition is required for a successful curvaton scenario. The curvaton acquires vacuum induced quantum fluctuations, which have amplitude
\begin{equation}
\label{amp}
\delta =\frac{H_{I}}{2\pi \phi_{I}}\,.
\end{equation}
These fluctuations are converted into the adiabatic density perturbations when the curvaton decays during its coherent oscillations or rotations. In order to match the observed amplitude of the fluctuations on the CMB, $r\delta\sim 10^{-5}$.

Let us first discuss the origin of the curvaton, which we take to be an $R$-parity conserving $D$-flat direction of the MSSM (for a review see~\cite{MSSM-rev}). Two candidate flat directions are ${\bf LLe}$ (where ${\bf L}$ denotes the left-handed slepton superfield and ${\bf e}$ the right-handed superfield) and $\mathbf{udd}$ (where $\mathbf{u}$ and $\mathbf{d}$ denote the right-handed squark superfields), which are lifted by the non-renormalizable operator:
\begin{equation}\label{supot1}
W\supset \frac{\lambda}{n}\frac{\Phi^{n}}{M_{\rm P}^{n-3}}\,,
\end{equation}
where $\lambda$ is a non-renormalizable coupling. For concreteness, we take the curvaton to be ${\bf LLe}$ so that the scalar component of the $\Phi$ superfield is:
\begin{equation}
\phi=({\widetilde L}+{\widetilde L}+{\widetilde e})/\sqrt{3},
 \end{equation}
  where
${\widetilde L}$ and ${\widetilde e}$ are the slepton and selectron scalar fields. At the lowest order the potential along the $\phi$ direction is given by:
\begin{equation}
V(\phi)=\frac{m_{\phi}^{2}|\phi|^{2}}{2} +\lambda^{2}\frac{|\phi|^{2(n-1)}}{M_{\rm P}^{2n-3}}+
\left(A\lambda\frac{\phi^{n}}{M_{\rm P}^{n-3}}+h.c.\right)\,,
\end{equation}
where $A\sim m_{\phi}\sim \mathcal{O}(100-1000)$~GeV, $m_{\phi}$ is the soft SUSY-breaking mass term, and $n=6$ for ${\bf LLe}$~\cite{MSSM-rev}. 

During inflation if $m^{2}_{\phi}\ll H^{2}_{I}$, the fluctuations along this nearly massless direction would create a homogeneous condensate with a vacuum expectation value (VEV) given by~\cite{MSSM-rev}
\begin{equation}
\phi_{I}\sim \left(m_{\phi}M_{\rm P}^{n-3}\right)^{1/n-2} \sim 10^{14}~\mathrm{GeV}\;,
\label{VEV}
\end{equation}
assuming $\lambda\sim\mathcal{O}(1)$. For $m_{\phi}\sim 100-1000$~GeV, and $n=6$, in order to match the amplitude of the density perturbations $\delta$, the Hubble expansion rate during inflation should be $H_{I}\sim 10^{10}$~GeV if $r\sim1$. 

There is a distinction between a positive and negative phase of the $A$ term. The difference in dynamics arises after the end of inflation. In the case of positive $A$ term the curvaton starts rolling towards the origin immediately, but in the case of a negative phase, for values of $A\geq \sqrt{40}m_{\phi}$, it may remain in a false vacuum with the VEV given by Eq. (\ref{VEV}). In this case the curvaton rotates instead of oscillates around its global minimum at $\phi=0$. In either scenario, the curvaton mass is negligible compared to the Hubble expansion rate. In fact, for $A=\sqrt{40} m_{\phi}$ and a negative phase the curvaton is actually massless along the real direction, and obtains inflaton-induced random fluctuations of order $\delta \phi\approx H_{I}/2\pi$. 

We now turn to the origin of $V_{0}$ within the MSSM. Let us consider a flat-direction orthogonal to the curvaton. If the curvaton is $\mathbf{LLe}$, this could be the ${\bf udd}$ direction. We take the inflaton direction to be: 
\begin{equation}
\sigma=({\widetilde u}+{\widetilde d}+{\widetilde d})/\sqrt{3}, 
\end{equation}
where ${\widetilde u}$ and ${\widetilde d}$ are squark scalars. Note that ${\bf udd}$ and ${\bf LLe}$ remain two  {\it independent} directions for the entire range of VEVs. 

This flat direction will also be lifted by the non-renormalizable operators. However, at larger VEVs the potential energy density stored in the ${\bf udd}$ direction will be larger than for the $\mathbf{LLe}$, so it would be lifted by higher order terms: 
\begin{equation}
\label{supot2}
W =\sum_{m\geq2}\frac{\lambda_{m}}{3m}\frac{\sigma^{3m}}{M_{\rm P}^{3m-3}}\,.
\end{equation}
The potential at lowest order would be:
\begin{equation}
\label{catpot}
V=\left |\lambda_{2}\frac{\sigma^{5}}{M_{\rm P}^{3}}+\lambda_{3}\frac{\sigma^{8}}{M_{\rm P}^{6}}+
\lambda_{4}\frac{\sigma^{11}}{M_{\rm P}^{9}}+\ldots \right |^{2}
\end{equation}
where $\ldots$ contain the higher order terms. Note that the $\lambda_m$ in  Eq.~(\ref{supot2}) are all non-renormalizable couplings induced by either gravity or by integrating out the heavy fields at the intermediate scale. At energies below the cut-off scale these coefficients need not necessarily be of $\mathcal{O}(1)$. 

Potentials like Eq.~(\ref{catpot}) were studied in Refs.~\cite{Dutta, Dutta-et.al.}. For $\lambda_{2}\ll \lambda_{3}\ll \lambda_{4}\ll \lambda_{n}\leq \mathcal{O}(1)$, they provide a unique solution for which first and second derivatives of the potential vanish along both radial and angular direction in the complex plane: $\partial V/\partial\sigma=\partial V/\partial \sigma^{\ast}=\partial^{2}V/\partial\sigma^{2}=\partial^{2}V/\partial\sigma^{\ast 2}=0$ (a saddle point condition)~\cite{Arnold}. For the first three terms in Eq.~(\ref{catpot}), it is possible to show that this happens when 
\begin{equation}\label{cond}
\lambda_{3}^{2}=\frac{55}{16} \lambda_{2} \lambda_{4}\,,
\end{equation}
at the VEVs: $\sigma=\sigma_{0} \exp{[i\pi/3,~i\pi,~i5\pi/3]}$, $\sigma_{0}=({2}/{11})({\lambda_{3}}/{\lambda_{4}})^{1/3}M_{\rm P}$. Concentrating on the real direction, the potential energy density stored in the inflaton sector is given by:
\begin{equation}
\label{relation}
 V_{0}\sim \left(\frac{153}{88}\right)^{2}\lambda_{2}^{2}\frac{\sigma_{0}^{10}}{M_{\rm P}^{6}}\,,
 \end{equation}
where $\sigma_{0}\ll M_{\rm P}$. Note that inflation occurs near the saddle point $\sigma_{0}$, where the effective mass vanishes. However, the third derivative of the potential is not negligible, $V^{\prime\prime\prime}\sim \lambda_{2}^{2}\sigma_{0}^{7}/M_{\rm P}^{6}\neq 0$, which leads to slow roll inflation. The potential for the inflaton becomes flat enough to sustain a large number of e-foldings.

As written, the condition Eq.~(\ref{cond}) represents a complete fine-tuning. Some deviation from this condition will be possible, changing the saddle point to a point of inflection, so long as $V^\prime$ remains small enough for sufficient e-foldings of inflation. Detailed discussion on fine-tuning in inflection point inflation can be found in Ref.~\cite{AEGM}.

The amplitude of perturbations of the inflaton is given by $\delta_{H}\sim V^{\prime\prime\prime}(\sigma_{0})N_{COBE}^{2}/30\pi H_{I}$~\cite{AEGM}.~The corresponding Hubble expansion rate is given by $H_{I}\sim (153/88)\lambda_{2}\sigma_{0}^{5}/M_{\rm P}^{4}$. For $\sigma_{0}\sim 10^{17.5}$~GeV and $\lambda_{2}\sim 10^{-6}$, it is possible to obtain $H_{I}\sim 10^{10}$~GeV, required for a successful curvaton scenario. For the above values, the inflaton perturbations are negligible, i.e. $\delta_{H}<10^{-5}$, therefore all the observed perturbations are created mainly by the decay of the curvaton. 

Now let us consider the aftermath of inflation. The inflaton would decay primarily into the MSSM {\it dof}. The coherent oscillations of the inflaton would give rise to instant preheating and thermalization of the light MSSM {\it dof} as discussed in Ref.~\cite{AFGM}, with a reheat temperature
\begin{equation}\label{Treh}
T_{R}\sim  [H_{I}M_{\rm P}]^{1/2}\sim 10^{13}~\mathrm{GeV}\,.
\end{equation}
However, not {\it all} of the MSSM {\it dof} will be in thermal equilibrium in our case. For the given choice of flat-direction fields, if both inflaton and curvaton simultaneously take large VEVs, the $SU(2)_{W}$ {\it dof} would not reach in thermal equilibrium, since the ${\bf LLe}$ VEV would induce large masses to those {\it dof}. This will play a crucial role in determining the non-Gaussianity parameter $f_\mathrm{NL}$, as we shall show below.

The curvaton $\phi$ starts to rotate about the origin when $H = H_\mathrm{osc}\sim m_{\phi}$. The field value at this time is $\vert\phi_\mathrm{osc}\vert\sim (m_{\phi}M_{\rm P}^{n-3})^{1/n-2}$. During this epoch the universe is already radiation-dominated following the decay of the inflaton. However, the curvaton cannot decay immediately, due to the fact that the curvaton VEV induces large masses $h\langle \phi(t)\rangle$ for gauge bosons, gauginos and (s)leptons, where $h$ is the gauge or Yukawa coupling. The curvaton's decay at leading order is {\it kinematically} forbidden if $h\langle \phi\rangle \geq  m_{\phi}/2\sim {\cal O}(\rm 100-1000)$~GeV. Decays do not occur until the Hubble expansion has redshifted $\langle \phi(t)\rangle $ down to $m_{\phi}/2h$.~Note that the SM Yukawa couplings are smaller than the gauge couplings.~Therefore the decays via SM Yukawas become kinematically allowed at higher VEVs.

During the rotations, the curvaton VEV will scale as $\phi(t)\propto a^{-3/2}$, as $a\propto H^{-1/2}$ during the radiation-dominated epoch. Therefore, each decay channel becomes allowed when~\cite{AD}
\begin{equation}
\label{latedec}
H=H_\mathrm{dec}\sim m_{\phi}\left({m_{\phi}}/{h\phi(t)}\right)^{4/3}\,,
\end{equation}
For large $\langle \phi(t)\rangle $, the decay time is naturally longer than the normal decay rate into the massless {\it dof}. The radiation energy density stored in the inflaton decay products scales as $\rho_{vis}\propto H^{2}$, where the subscript denotes the visible {\it dof}. The ratio of the energy densities at the time the curvaton decays is given by
\begin{eqnarray}
r &\equiv& \frac{\rho_{\phi}}{\rho_\mathrm{vis}} \sim \left.\frac{\rho_{\phi}}{\rho_\mathrm{vis}}\right |_\mathrm{osc}\left(\frac{H_\mathrm{dec}}{H_\mathrm{osc}}\right)^{-1/2}\,, \nonumber \\
&\sim &\left(\frac{m_{\phi}}{M_{\rm P}}\right)^{2/(n-2)}\left(\frac{m_{\phi}}{h\phi}\right)^{-2/3} \leq 1\,.
\end{eqnarray}
The kinematical blocking due to the curvaton VEV enhances the efficiency factor, $r$, therefore the curvaton rotations prolong the mater-dominated epoch till it decays completely. For soft SUSY-breaking mass $m_{\phi}\lesssim 1$~TeV, the inefficiency parameter is $r\sim\mathcal{O}(1)h^{2/3}$. Although the LHC has already placed severe constraints on the parameter space for low-scale SUSY, the current limits do not exclude heavy squark and slepton masses $\gtrsim500$~GeV~\cite{Atlas}. Since our flat directions are all made up of squarks and sleptons, there is a large parameter space available in which this condition may be satisfied if SUSY is discovered at the LHC.

Since the curvaton decay is delayed due to the kinematical blocking, $r\leq1$ is different for each decay channel. What range of $f_\mathrm{NL}$ we expect from the various dominant decay channels of the curvaton depends on the different values of $h$. If we consider the SM gauge couplings, then $h\sim 0.1$ and we would expect the largest $f_\mathrm{NL}\sim (5/4r)\sim {\cal O}(1)h^{-2/3}\sim {\cal O}(5)$. However, the curvaton also has the Yukawa interactions, especially when the curvaton decays into leptons and sleptons, for which:
\begin{equation}\label{f-NL}
f_\mathrm{NL}\sim \frac{5}{4r} \sim {\cal O}(1) h^{-2/3}\sim 10 -10^{3}\,,
\end{equation}
for $h\sim 10^{-2} -10^{-5}$. This range of $h$ covers all the SM Yukawas except the top Yukawa which is of order $h\sim 0.1$. Due to the smaller values of $h$, these decays are kinematically allowed at higher VEVs. An exact prediction for net effect on $f_\mathrm{NL}$ requires a complete analysis of the decay modes for the ${\bf LLe}$ curvaton which is beyond the scope of the current letter, but it can be seen that this model of the curvaton can provide $f_\mathrm{NL}$ in a range which will be observationally relevant in the near future.

The temperature at which the curvaton decay products reach thermal equilibrium is determined by Eq.~(\ref{latedec}). The final thermal bath filled with MSSM {\it dof} would be obtained by the reheat temperature
\begin{equation}
\label{finalT}
T_{R,f}\sim (H_\mathrm{dec}M_{\rm P})^{1/2}\sim 10^{4.5}-10^{6.5}~\mathrm{GeV}
\end{equation}
for $h\sim 10^{-2}-10^{-5}$. Such a temperature is sufficient to excite weakly interacting massive particles and for baryogenesis~\cite{RMP}. Note that both the temperatures from Eqs.~(\ref{Treh}) and (\ref{finalT}) are sufficiently high to excite thermal/non-thermal gravitinos and axinos. If the gravitinos or axinos are the lightest SUSY particle, this causes two problems for this scenario: over-production of gravitinos with both helicities would be bad for BBN, and the gravitinos and axinos would thermally decouple even before the curvaton has started decaying. This would generate large residual isocurvature perturbations, because gravitinos and axinos can never come into thermal equilibrium. Instead the ideal dark matter candidate would be the neutralino, which decouples from the thermal plasma at $T\sim 40-50$~GeV.

Our discussion so far has been based on treating ${\bf udd}$ as the inflaton and ${\bf LLe}$ as the curvaton flat direction. In principle, we could have swapped the roles of inflaton and curvaton, i.e. ${\bf LLe}$ as an inflaton and ${\bf udd}$ to be the curvaton. The main aspects of the analysis would not differ at all. Although treating ${\bf udd}$ as a curvaton would also make $SU(3)_{c}$ {\it dof} heavy during the curvaton oscillations and this would alter the detailed discussion of thermalization, nevertheless the range of $f_\mathrm{NL}$ quoted above for the SM Yukawas in Eq.~(\ref{f-NL}) would remain the same.

To summarize, we have discussed the possibility of constructing a model in which both the inflaton and curvaton are flat direction fields within the MSSM. The radiation created from the decay of the inflaton and curvaton belongs to the visible sector, avoiding the problem of residual isocurvature fluctuations, while the curvaton mechanism can create observable non-Gaussianity. The non-Gaussianity parameter $f_\mathrm{NL}$ depends crucially on the SM gauge and Yukawa couplings, and ranges from $\mathcal{O}(5)$ to $\mathcal{O}(1000)$ in the different decay channels (for Yukawas in the range $h\sim 10^{-2}-10^{-5}$, which is the case for all the Yukawas except the top). The model favours a visible-sector dark matter candidate such as the lightest neutralino but will not work if the lightest SUSY particle is a gravitino or axino type.

{\it Acknowledgements:} The authors would like to thank R. Allahverdi, P. Dayal, A. Liddle, and  D. Wands for helpful discussions.


\end{document}